\documentclass{webofc}
\def\pt{\mbox{$p_{\rm T} $}}  
\usepackage[varg]{txfonts}   
\usepackage{color}
\usepackage[colorlinks=true]{hyperref}
\wocname{EPJ Web of Conferences}
\woctitle{Resonance Workshop at Catania}
\begin{document}
\title{How much does the hadronic phase contribute to the observed anisotropic flow at the LHC?}
%
%

\author{Raimond Snellings\inst{1,2}\fnsep\thanks{\email{R.J.M.Snellings@uu.nl}} }

\institute{
Nikhef National Institute for Subatomic Physics, Amsterdam, The Netherlands
\and 
Utrecht University, Princetonplein 5, 3584 CC Utrecht, The Netherlands.
}

\abstract{%
Elliptic flow signals the presence of multiple interactions between the constituents of the created matter in heavy-ion collisions. 
This includes possible contributions from the different phases, including the hadronic phase. 
In these proceedings I will first show that the energy dependence of elliptic flow, based on recent ALICE and STAR beam 
energy scan measurements, can largely be understood in terms of a boosted thermal system. In addition, 
a detailed comparison between the identified particle elliptic flow measured by the ALICE collaboration and viscous hydrodynamical 
model calculations with and without a hadronic afterburner is performed to constrain the possible effects 
of individual hadron-hadron re-interaction cross-sections. 
  }
\maketitle
%

\section{Introduction}

In the world around us, quarks and gluons do not exist as free particles because they are permanently bound 
into hadrons by the strong interaction. 
At very high temperatures and densities however, hadronic matter is expected to undergo a phase transition to 
a new state of matter, the Quark-Gluon Plasma (QGP), where quark and gluon degrees of freedom are not 
anymore confined inside the hadrons. 
Collisions of heavy ions at RHIC and at the LHC energies produce a system with temperatures that 
are well above the strong phase transition temperature and are therefore a unique tool to create 
and study the complex properties of the QGP and the QCD phase transition to ordinary hadronic matter in the laboratory.

The experimental evidence that the QGP properties resemble those of a strongly interacting 
liquid has led to a consensus that much of the dynamical evolution of the created system can be 
modelled using viscous relativistic hydrodynamics.   
\begin{figure}[thb]
\begin{center}
\includegraphics[width=12cm]{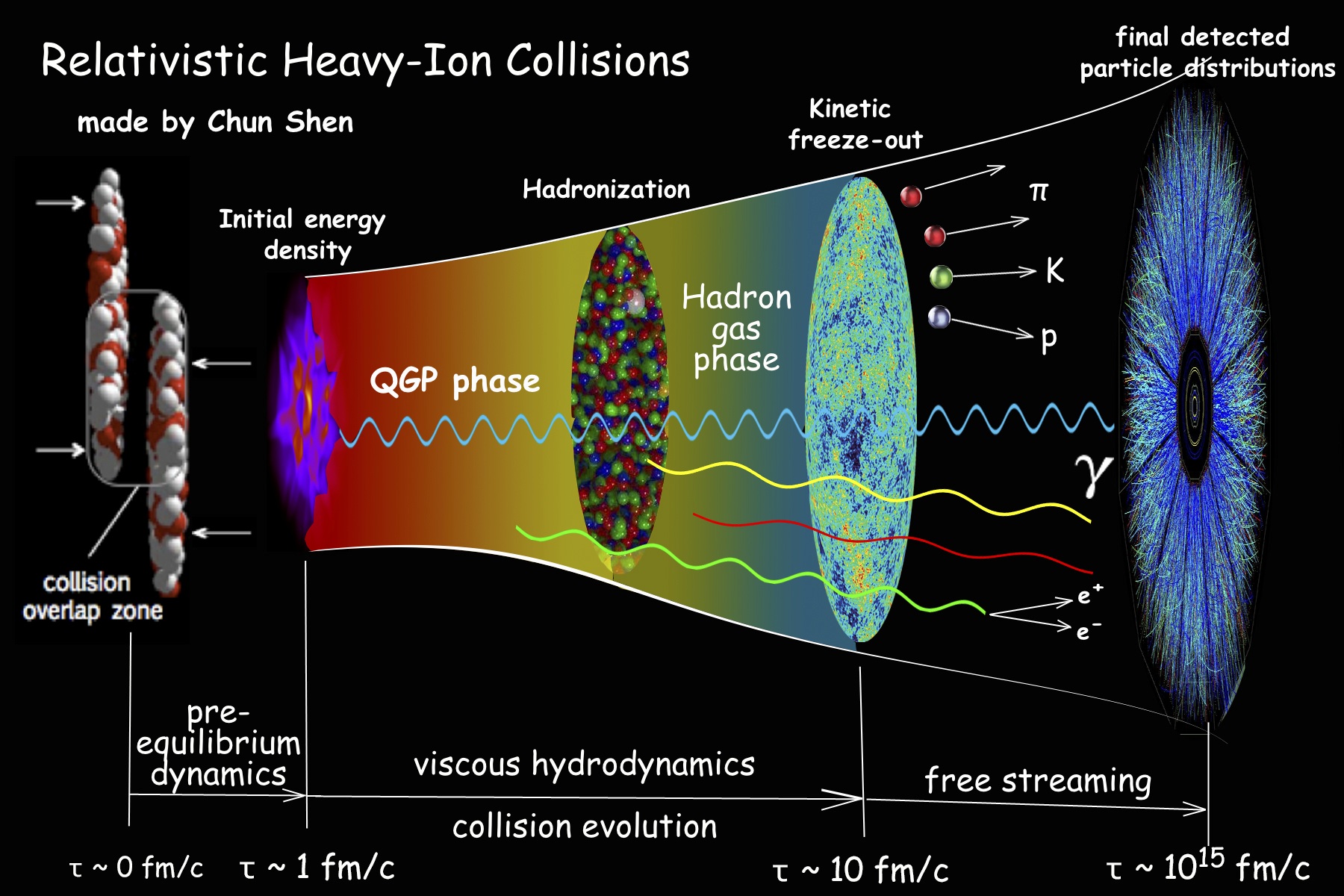}
\end{center}
\caption{
Cartoon of the time evolution of an ultra-relativistic heavy-ion collision. 
Figure taken from~\cite{sketch}.
}
\label{fig:figure1} 
\end{figure}
Figure~\ref{fig:figure1} shows a cartoon of the time evolution of the collision and shows 
that after a short pre-equilibrium phase the system evolution is modelled in terms of 
viscous hydrodynamics.

It is observed that the bulk hadron yields in heavy-ion collisions are well described 
in thermal and statistical models assuming chemical equilibrium. 
These models successfully describe almost all bulk hadron yields, from center of mass 
energies of a few GeV to a few TeV~\cite{Stachel:2013zma}. 
In these models only two parameters, the chemical freeze-out temperature $T_{\rm chem}$ and 
baryochemical potential $\mu_B$, are most relevant.
The $T_{\rm chem}$ extracted from thermal and statistical model fits to the 
data is close to the phase transition temperature obtained from recent lattice calculations and is found to 
saturate from RHIC up to LHC energies, where heavy-ions collide at an order of 
magnitude higher center of mass energy. 
In addition, measurements show that $T_{\rm chem}$ is also constant 
as a function of collision centrality, which is naturally expected in these models.

Nevertheless, clear deviations from thermal and statistical model calculations are observed at the LHC 
in e.g. the changing $K^{*0}/K^{-}$ ratio as function of centrality~\cite{Abelev:2014uua}  and in the proton and 
anti-proton yields~\cite{Abelev:2013vea} (although in this case the deviations are less than 3$\sigma$). 
For the success of the thermal and statistical model description this is a fly in the ointment and 
these deviations of the proton and anti-proton yields were part of 
a larger so-called {\it proton puzzle} at the LHC 
(at lower energies similar deviations are observed~\cite{Becattini:2012sq}).

A possible explanation for these deviations could be that statistical hadronization occurs 
out of chemical equilibrium~\cite{Begun:2014rsa}, while another scenario could be modifications due to 
some inelastic hadron-hadron final stage interactions (e.g. baryon-antibaryon annihilation~\cite{Becattini:2012xb}) 
at temperatures below $T_{\rm chem}$. 

In contrast to the bulk hadron yields, the momentum distributions of the finally emitted hadrons 
can not be described as 
a thermal system with $T_{\rm chem}$ but instead as a boosted thermal system, 
with a kinetic freeze-out temperature $T_{\rm kin}$ and radial flow 
velocity $\beta_0$. It is found that the observed $T_{\rm kin}$ is smaller than $T_{\rm chem}$, 
which is considered evidence for quasi-elastic resonance scattering in the hadronic phase. 

If the created system stays close to kinetic equilibrium during the hadronic phase 
it can be modelled using viscous hydrodynamics, 
while accounting for chemical freeze-out at $T_{\rm chem}$. 
An example of such a model is {\sc VISH2+1}~\cite{Song:2007fn,Song:2008si}. 
However, if the system becomes 
too dissipative in the hadronic phase viscous hydrodynamics breaks down, and in this case a microscopic 
hadron cascade model should provide a more reliable description. 
{\sc VISHNU}~\cite{Song:2010aq} is an example of a model which matches a viscous 
hydrodynamical model {\sc VISH2+1} to 
a hadron cascade model {\sc UrQMD}~\cite{Bass:1998ca} for the dissipative hadronic phase. 

In the following sections of these proceedings I will first show that the integrated and $\pt$-differential 
elliptic flow for charged and identified particles matches globally our expectations from a boosted 
thermal system. However, a detailed comparison of identified particle elliptic flow measured 
by the ALICE collaboration with viscous hydrodynamical calculations shows that 
a quantitative comparison fails for more central collisions. 
Including the contributions from individual hadron-hadron re-interaction cross-sections, 
by augmenting the viscous hydrodynamical 
calculations using a hadron cascade model, significantly 
modifies the $\pt$-differential elliptic flow compared to pure viscous hydrodynamical calculations. 
However this currently does not seem to improve the description of the data.
Finally, because resonances can provide more differential 
information on the hadronic processes, I will present in the last section the $\pt$-differential 
$\phi$-meson elliptic flow.

\section{Charged Particle Elliptic Flow}

Anisotropic flow is an important observable in ultra-relativistic heavy-ion collisions
as it signals the presence of multiple interactions between the constituents of the created matter.
This includes possible contributions from the different phases including the hadronic phase. 
Therefore, anisotropic flow has been observed in nucleus--nucleus collisions from 
low energies up to $\sqrt{s_{_{\rm NN}}} = 2.76$~TeV at the Large Hadron Collider 
(LHC)~\cite{Voloshin:2008dg,Heinz:2013th,Aamodt:2010pa}. 
The azimuthal anisotropic flow is usually characterized by the Fourier coefficients~\cite{Voloshin:1994mz}:
\begin{equation}
v_n = \langle \cos [n(\phi - \Psi_n)]\rangle,
\label{eq:harmonics}
\end{equation}
where $\phi$ is the azimuthal angle of the particle, $\Psi_n$ is the azimuthal angle of the plane of symmetry, 
and $n$ is the order of the harmonic. The second Fourier coefficient $v_2$ is called elliptic flow~\cite{Ollitrault:1992bk}.

\begin{figure}[th]
\includegraphics[width=7.4cm]{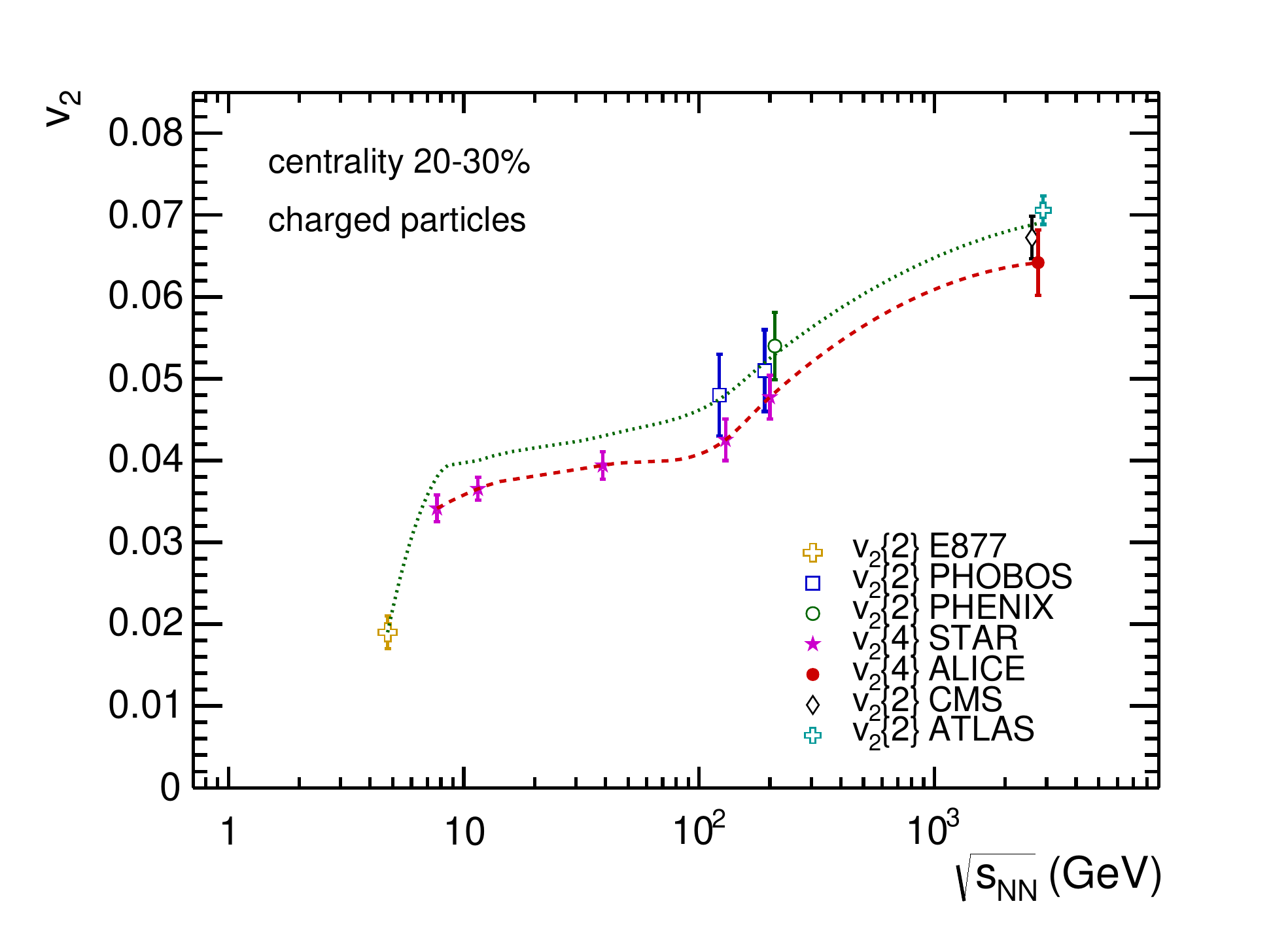}
\includegraphics[width=7.4cm]{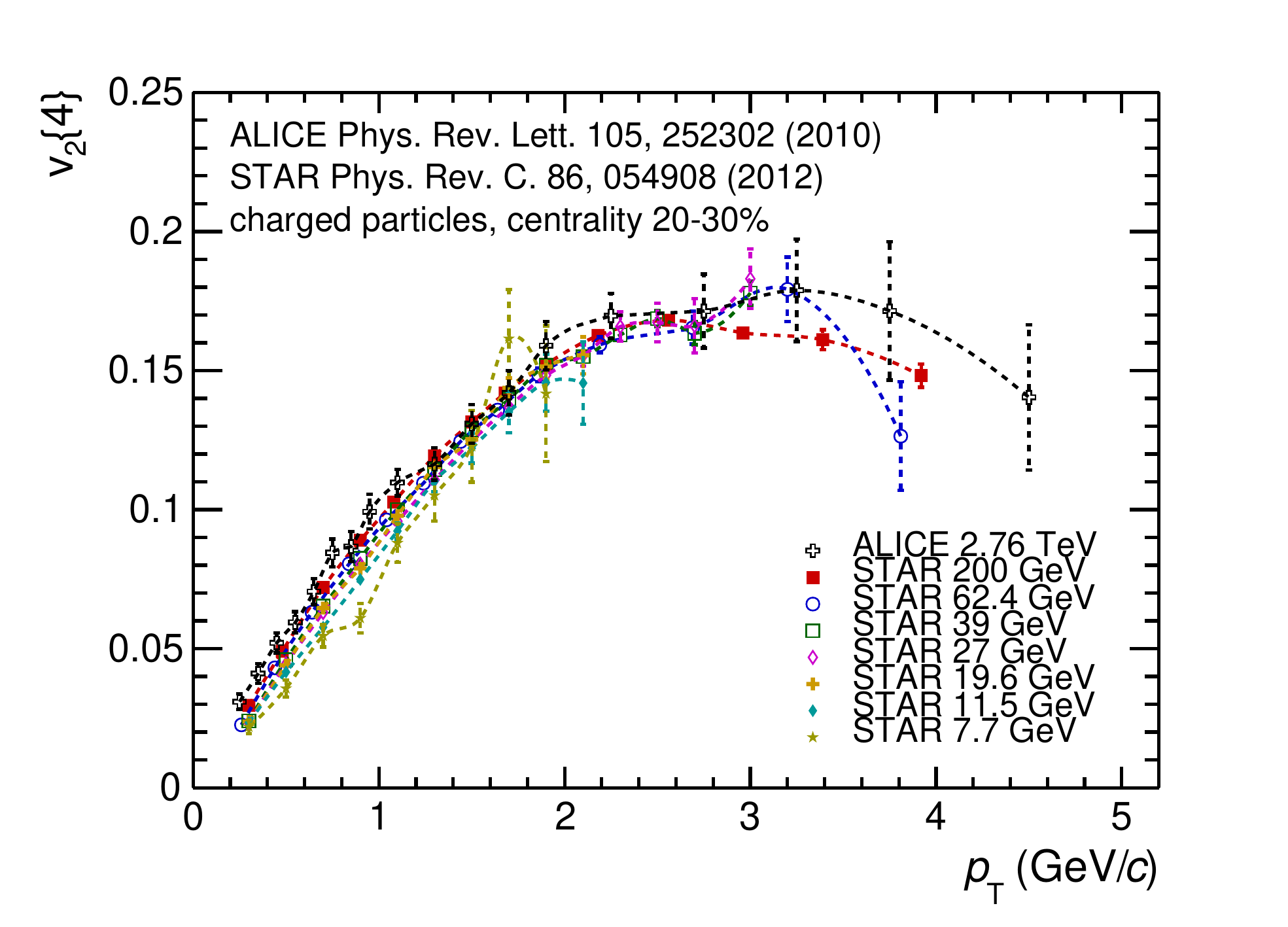}
\caption{
Left: Elliptic flow for charged particles as function of center of mass energy.
Right: The charged particle $\pt$-differential elliptic flow as function of center of 
mass energy~\cite{Aamodt:2010pa,Shi:2012ba,Adamczyk:2012ku}.
}
\label{fig:figure2} 
\end{figure}

In the left panel of Fig.~\ref{fig:figure2} the elliptic flow as function of center of  mass energy is 
plotted for charged hadrons.
Experimentally, because the planes of symmetry $\Psi_n$ in Eq.~\ref{eq:harmonics} are not known, 
the anisotropic flow coefficients are estimated from measured correlations between the 
observed particles. 
Here the elliptic flow estimated from two- and four-particle correlations is denoted by $v_2$\{2\} 
and $v_2$\{4\}, respectively. 
The $v_2$ obtained from two-particle correlations  
is in general not completely due to collective effects, it also contains contributions from 
so-called non-flow (these are other sources of azimuthal correlations for instance due to 
jets and resonance decays). Experimentally they are suppressed as much as possible using  
kinematic cuts between the correlated particles. 
For the remaining significant difference observed between flow estimates from 
two- and four-particle correlations at LHC and RHIC energies, we have strong evidence 
that this is mainly due to event-by-event fluctuations in the elliptic flow.  

The figure shows an interesting center of mass dependence of the elliptic flow which, with the recent addition of the LHC and RHIC beam energy scan results, provides tantalising evidence for a slow increase of the elliptic flow between 7.7 and 130 GeV followed by a much steeper increase from the 
top RHIC to LHC energy. This behaviour is similar to what has been observed in the slopes of the particle 
spectra and resembles predictions by L. van Hove for the $\langle \pt \rangle$~\cite{vanhoven}.
 
The right panel of Fig.~\ref{fig:figure2} shows the $\pt$-differential elliptic flow of charged particles 
for center of mass energies ranging from 7.7 GeV to 2.76 TeV. Carefully 
examining the data, an increase of the $\pt$-differential $v_2\{4\}$ is observed with 
increasing center of mass energy, however this increase is small. 
This shows that a significant fraction of the increasing elliptic flow is due to an 
increase in the $\langle \pt \rangle$~\cite{Abelev:2013bla}. 
A natural question to ask is if this is because a boosted thermal system is created for which the 
collective flow, and therefore the $\langle \pt \rangle$, increases with center of mass energy. 
To answer this question we examine in the next section the $\pt$-differential elliptic flow in 
much more detail with measurements of particles with very different masses.
 
\section{Identified Particle Elliptic Flow}

The modification to the  $\pt$-differential $v_2$ due to an increase in collective flow, 
i.e. both the radial flow, $\beta_{0}$, and its  azimuthal variation, $\beta_2$, 
with increasing center of mass energy can be illustrated using a so-called 
blast-wave parameterisation.
While the blast-wave model is simple and does not contain any dynamical information, it 
can be used to describe main features of the spectra, anisotropic flow and femtoscopy 
with a few parameters. 
The main model parameters are the 
kinetic freeze-out temperature, $T_{\rm kin}$, the radial flow, its azimuthal variation and,  
the source density at freeze-out~\cite{Adler:2001nb,Retiere:2003kf}.  

\begin{figure}[th]
\includegraphics[width=7.4cm]{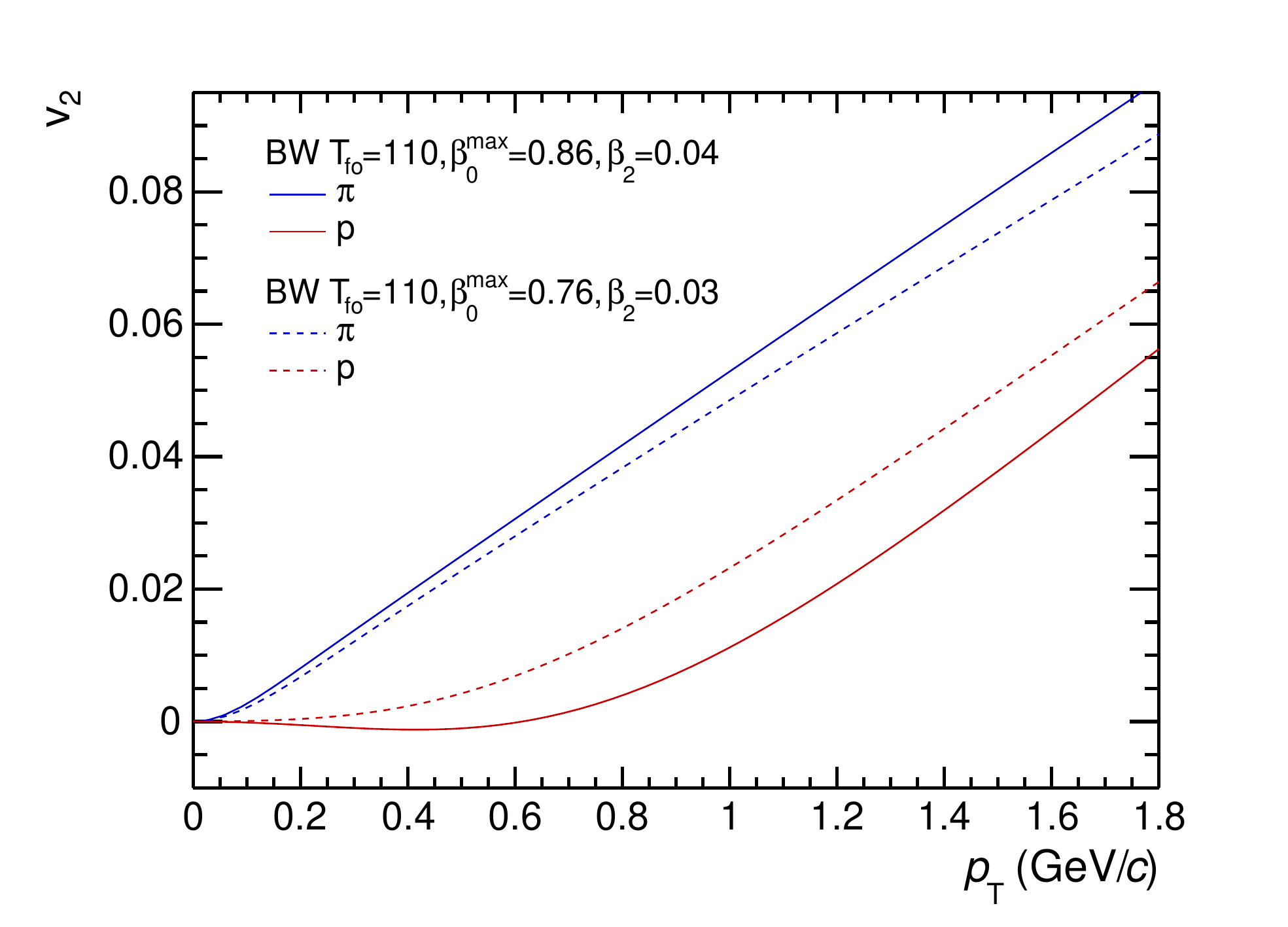}
\includegraphics[width=7.4cm]{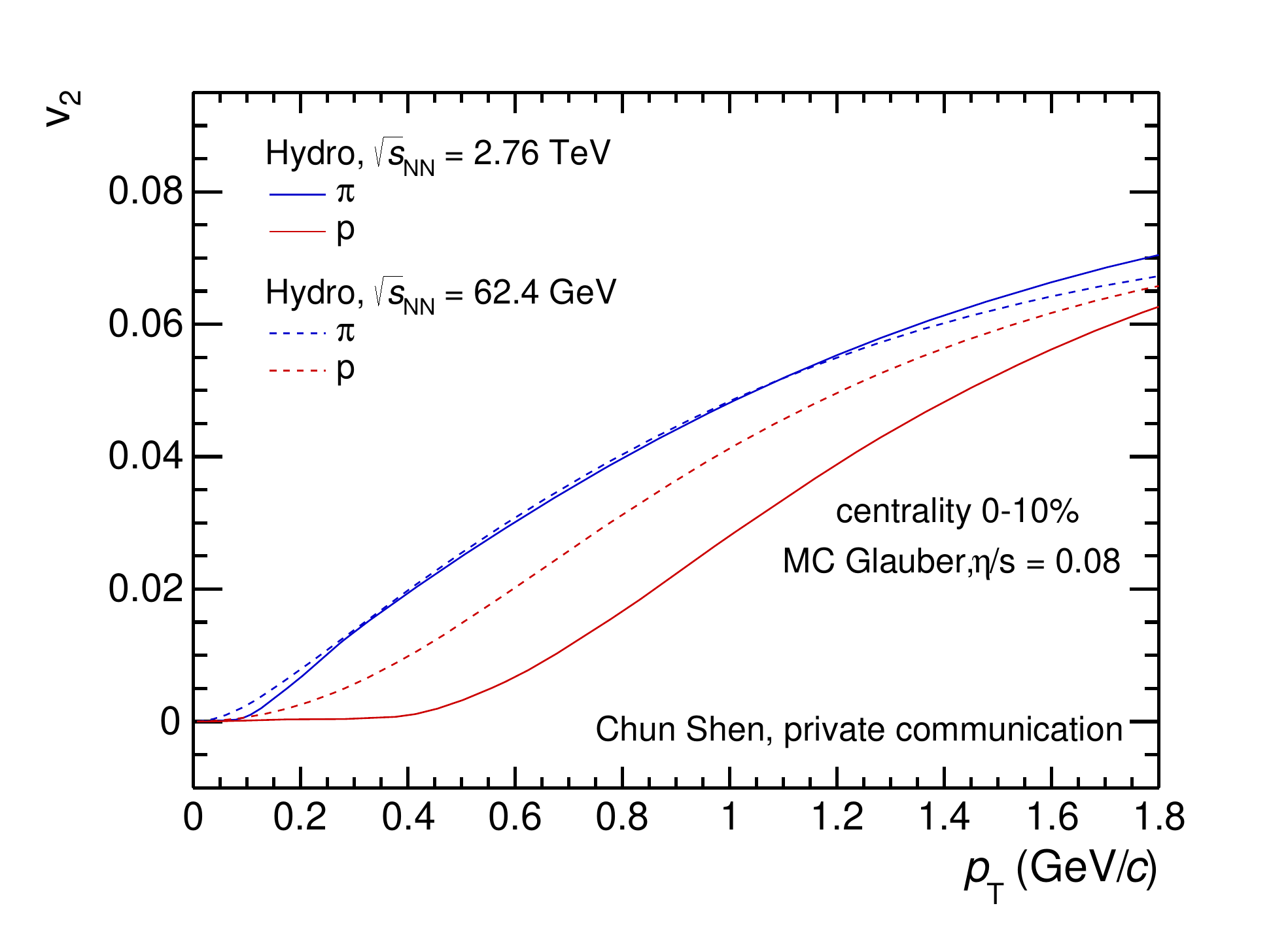}
\caption{
Left: Blast-wave results for identified particle $\pt$-differential $v_2$ for two systems with 
different collective flow.
Right: Viscous hydrodynamical calculations of identified particle $\pt$-differential $v_2$ at
two collision energies.
}
\label{fig:figure3} 
\end{figure}
Fig.~\ref{fig:figure3} (left) shows the $\pt$-differential $v_2$ from the blast-wave model for two 
different magnitudes of the collective flow and its azimuthal variation with all other parameters fixed. 
The plot clearly shows a characteristic mass splitting in the $\pt$-differential $v_2$ for pions and protons. 
The mass splitting originates from an interplay between the blueshift generated by the radial flow 
and its azimuthal variation. 
At low-$\pt$ we can express the $\pt \approx \pt^{\rm th} + mc\beta$ as the sum of a 
thermal contribution (independent of the mass of the hadron) and a radial flow component 
with flow velocity $\beta$ (this part is proportional to the mass of the hadron). 
The blueshift generates a mass dependent modification to the hadron $\pt$-spectra and 
determines the magnitude of the mass splitting in the $\pt$-differential $v_2$ of the hadrons. 
The blast-wave curves indeed show that the mass splitting increases from the 
dashed curves (smaller radial flow) to the solid curves (larger radial flow). 

In relativistic viscous hydrodynamics, a dynamical model of the system evolution, the 
$\pt$-differential $v_2$ depends on many more parameters, e.g. initial conditions, 
equation of state, transport parameters such as $\eta/s$ and freeze-out conditions. 
At different center of mass energies, some of these parameters are typically constrained  
by the measured integrated charged hadron yields and pion spectra in central collisions. 
As a result the centrality and mass dependence of the $\pt$-differential $v_2$ can be used 
to check the validity of the model and to constrain the main transport parameters. 
More recently, measurements of the event-by-event fluctuations in $v_2$ and additional higher 
harmonics, such as $v_3$, are used to provide stringent constraints on the initial conditions. 

In the right panel of fig.~\ref{fig:figure3} the $\pt$-differential $v_2$  for pions and protons is plotted from viscous 
hydrodynamical calculations, using MC Glauber initial conditions and a kinetic viscosity $\eta/s = 0.08$. 
These detailed model calculations also clearly show the characteristic mass splitting, which increases with 
increasing center of mass collision energy (dashed and solid curves are examples for RHIC and LHC 
energies, respectively). 

\begin{figure}[th]
\includegraphics[width=7.4cm]{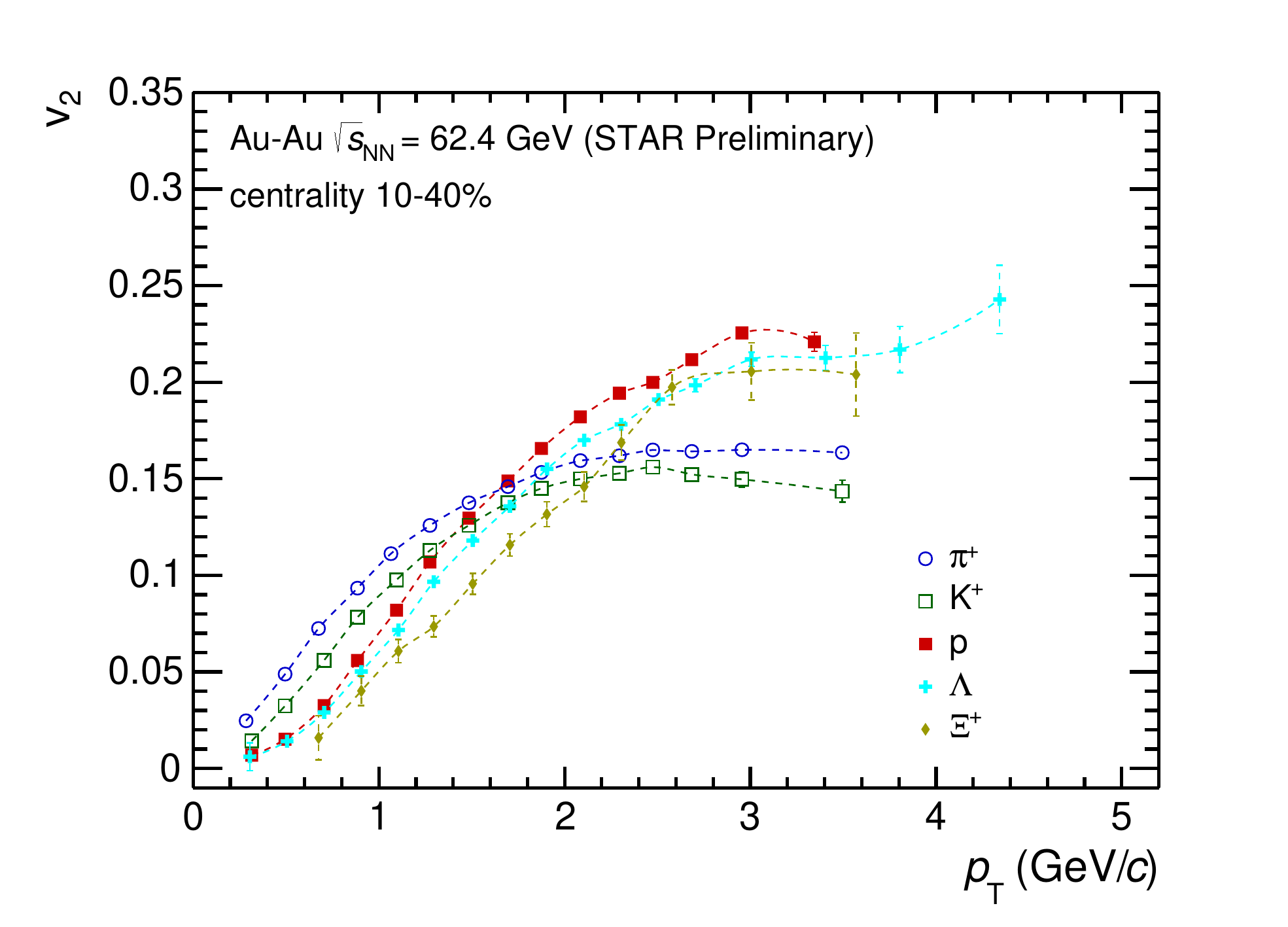}
\includegraphics[width=7.4cm]{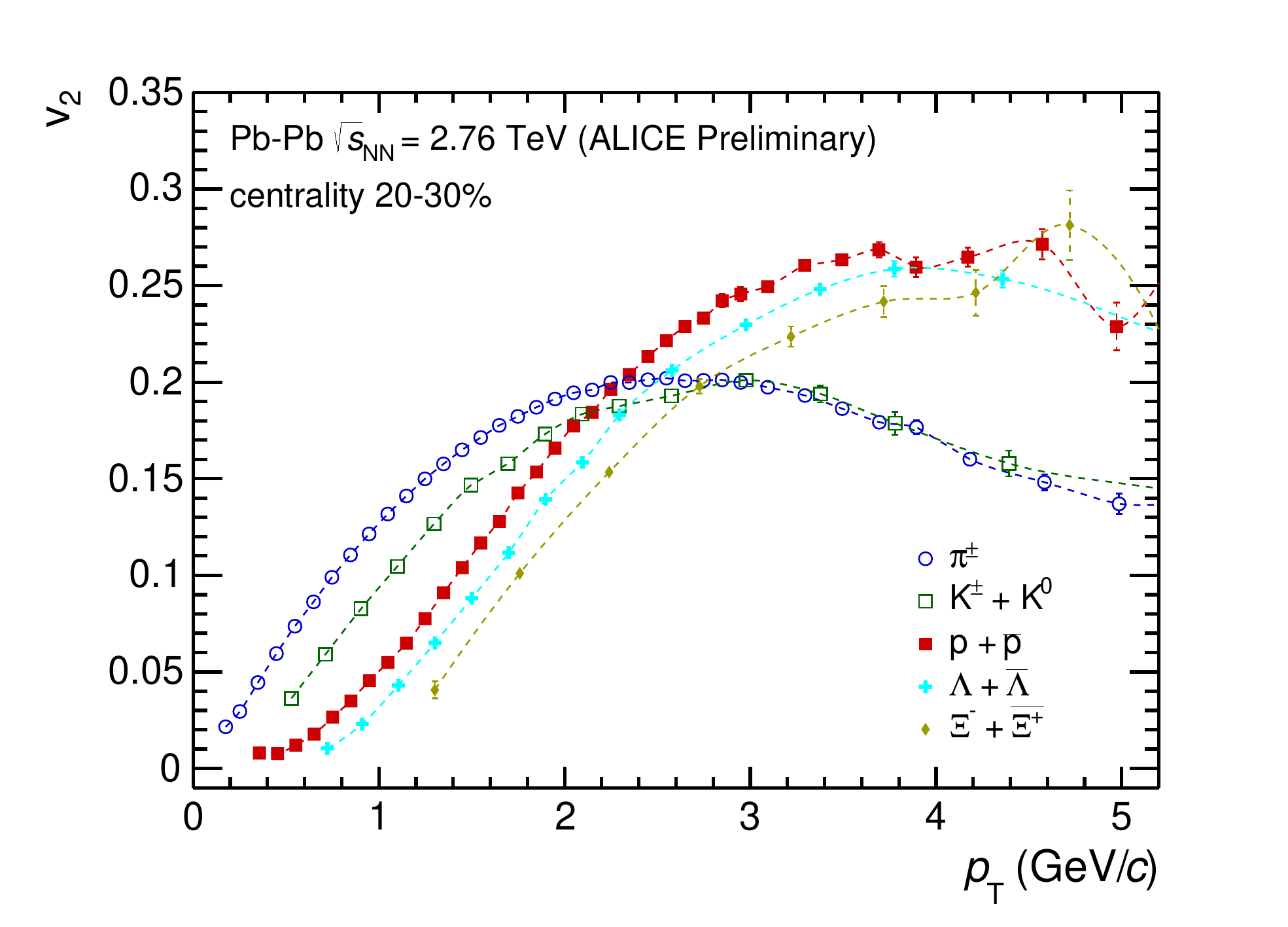}
\caption{
Left: STAR $\pt$-differential identified particle $v_2$ for Au--Au collisions at $\sqrt{s_{_{\rm NN}}} = 62.4$ GeV.
Right: ALICE $\pt$-differential identified particle $v_2$ for Pb--Pb collisions at 
$\sqrt{s_{_{\rm NN}}} = 2.76$ TeV~\cite{Abelev:2014pua} 
}
\label{fig:figure4} 
\end{figure}
Recently the STAR collaboration has performed anisotropic flow measurements in a range of center of mass energies 
as part of their beam energy scan program. In Fig.~\ref{fig:figure4} (left) the $\pt$-differential $v_2$ is shown for 
pions, kaons, protons, lambda and cascade, measured at $\sqrt{s_{_{\rm NN}}} = 62.4$ GeV.  
For $\pt < 1.5$ GeV/$c$ a clear mass ordering is observed for the particles which vary in mass by an order of magnitude.
In the right panel of Fig.~\ref{fig:figure4} the measurements at $\sqrt{s_{_{\rm NN}}} = 2.76$ TeV, by the ALICE collaboration, are plotted. 
The same mass ordering, for $\pt < 2$ GeV/$c$, is again 
observed, however we see that the mass splitting is much more pronounced at a larger center of mass energy.

Ideal hydrodynamics predicts that the mass splitting pattern persists even at large values of $\pt$ in contrast to what is observed in the data at RHIC and the LHC, this is apparent from Fig.~\ref{fig:figure4} above $\pt \approx$~2~GeV/$c$. 
An elegant explanation of the particle type dependence and magnitude of $v_2$ at larger $\pt$ 
is provided by the coalescence picture~\cite{Molnar:2003ff}.

\begin{figure}[th]
\includegraphics[width=7.4cm]{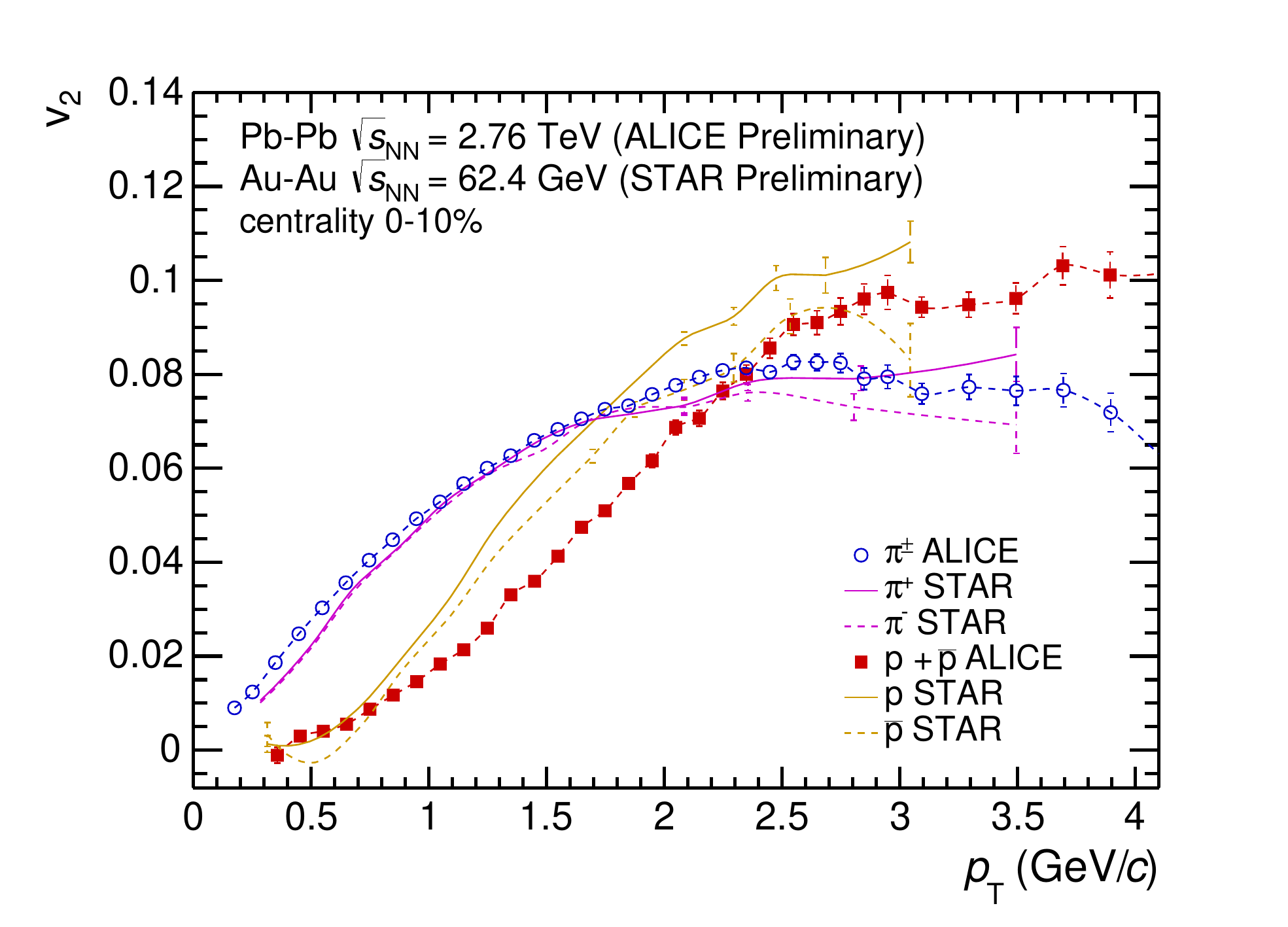}
\includegraphics[width=7.4cm]{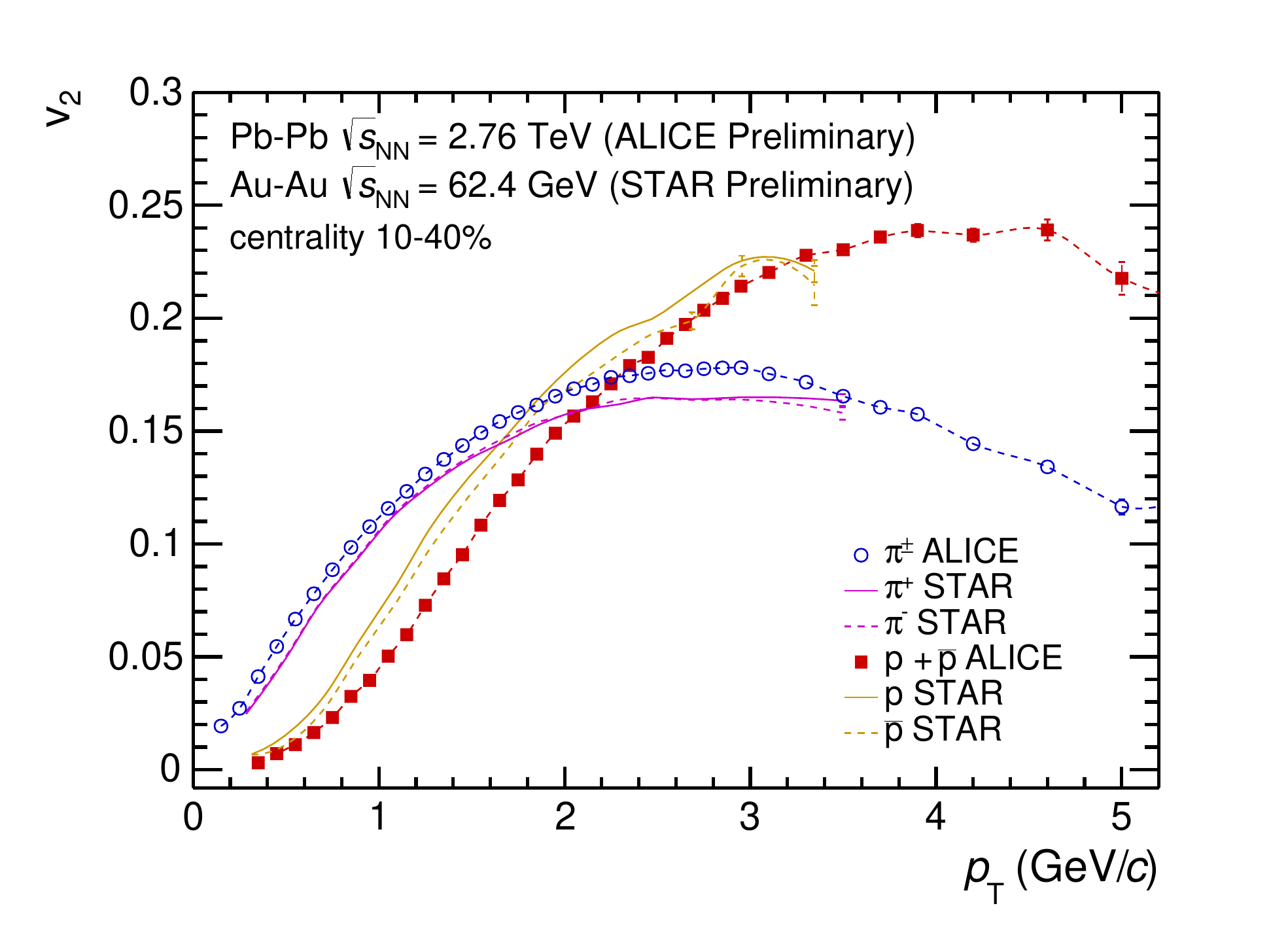}
\caption{
Comparison of $\pt$-differential $v_2$ at $\sqrt{s_{_{\rm NN}}} = 62.4$~GeV~\cite{Adamczyk:2013gw} 
and 2.76 TeV~\cite{Abelev:2014pua} for different particle species. 
Left: for the central 0--10\% collisions.
Right: for the collisions in the 10--40\% centrality percentile.
}
\label{fig:figure5} 
\end{figure}
To compare the collision energy dependence of the mass splitting we compare for two centralities the 
$\pt$-differential $v_2$ for pions and protons measured by STAR and ALICE. 
Figure~\ref{fig:figure5} (left) shows the comparison for the 
0--10\% centrality percentile. Because at RHIC energies we observe a small difference between the 
particle and antiparticle anisotropic flow~\cite{Adamczyk:2013gv}, the STAR results are plotted 
separately for $\pi^{+}$, $\pi^{-}$ and $p$, $\bar{p}$. 
In the right panel of Fig.~\ref{fig:figure5} the same comparison is shown for more peripheral, 10--40\%, collisions. 
This comparison clearly illustrates the significant increase of the mass splitting between pions and protons 
as function of collision energy,  for both centrality ranges.
Such an increase is expected if the collective flow increases with collision energy and is therefore 
in qualitative agreement with hydrodynamical model calculations.

\begin{figure}[th]
\includegraphics[width=7.4cm]{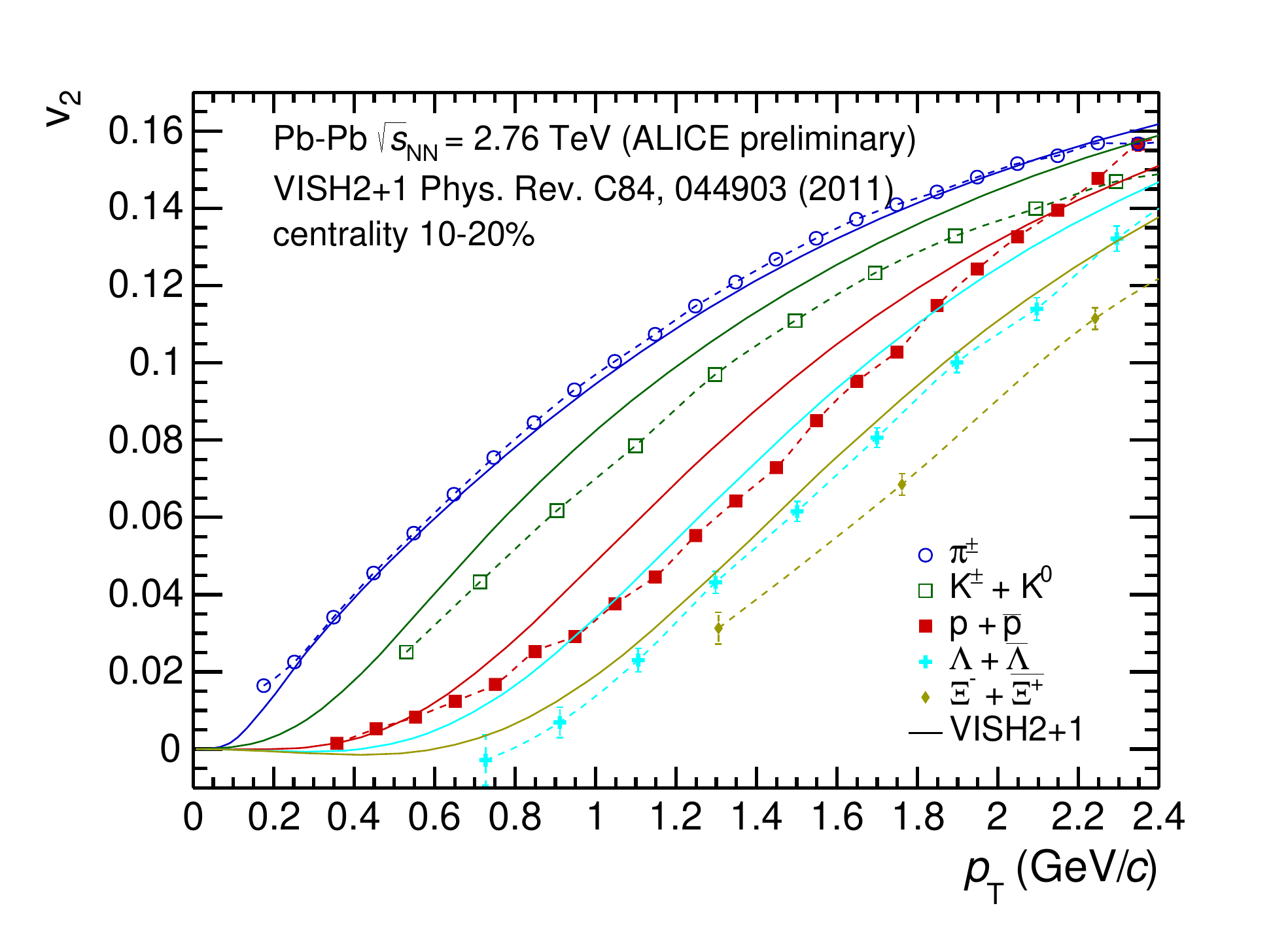}
\includegraphics[width=7.4cm]{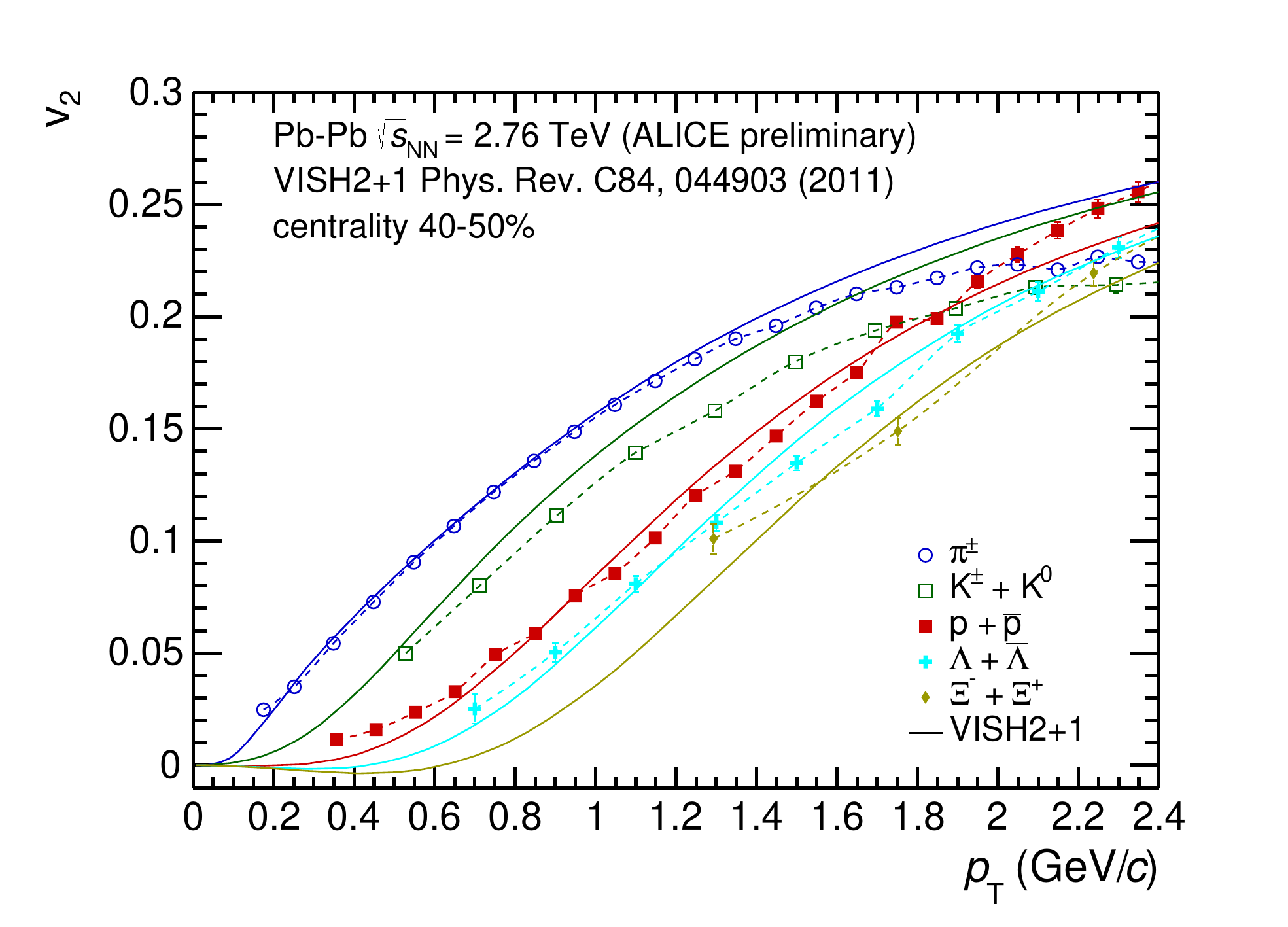}
\caption{
Comparison between ALICE measurements of $\pt$-differential $v_2$ at 
$\sqrt{s_{_{\rm NN}}} = 2.76$ TeV~\cite{Abelev:2014pua} 
and viscous hydrodynamical model predictions. 
Left: for the central 10--20\% collisions.
Right: for the collisions in the range of 40--50\% centrality percentile. The dashed lines connect the datapoints.
}
\label{fig:figure6} 
\end{figure}
First preliminary measurements of the $\pt$-differential $v_2$ at $\sqrt{s_{_{\rm NN}}} = 2.76$ TeV for 
pions, kaons and protons were presented and compared to viscous hydrodynamical 
model predictions at QM2011~\cite{Krzewicki:2011ee}. 
For mid-central collisions, 40--50\%, a fair agreement between the 
model calculations and the data was observed. In Fig.~\ref{fig:figure6} the comparison  
between viscous hydrodynamical calculations and the data is shown, here also including the 
lambdas and cascades. 
Also for the lambdas and cascades there is a fair agreement between data and theory.

However, for more central collisions a clear discrepancy between data and theory was 
observed for the protons. 
This discrepancy, also observed in the yields and spectra of the protons and anti-protons, is part of what is called  
the {\it proton puzzle} at the LHC. Recent ALICE measurements of the 
$\pt$-differential $v_2$, which now include more particles, show that this discrepancy is not only 
observed for the protons but also for other baryons.

\begin{figure}[th]
\includegraphics[width=7.4cm]{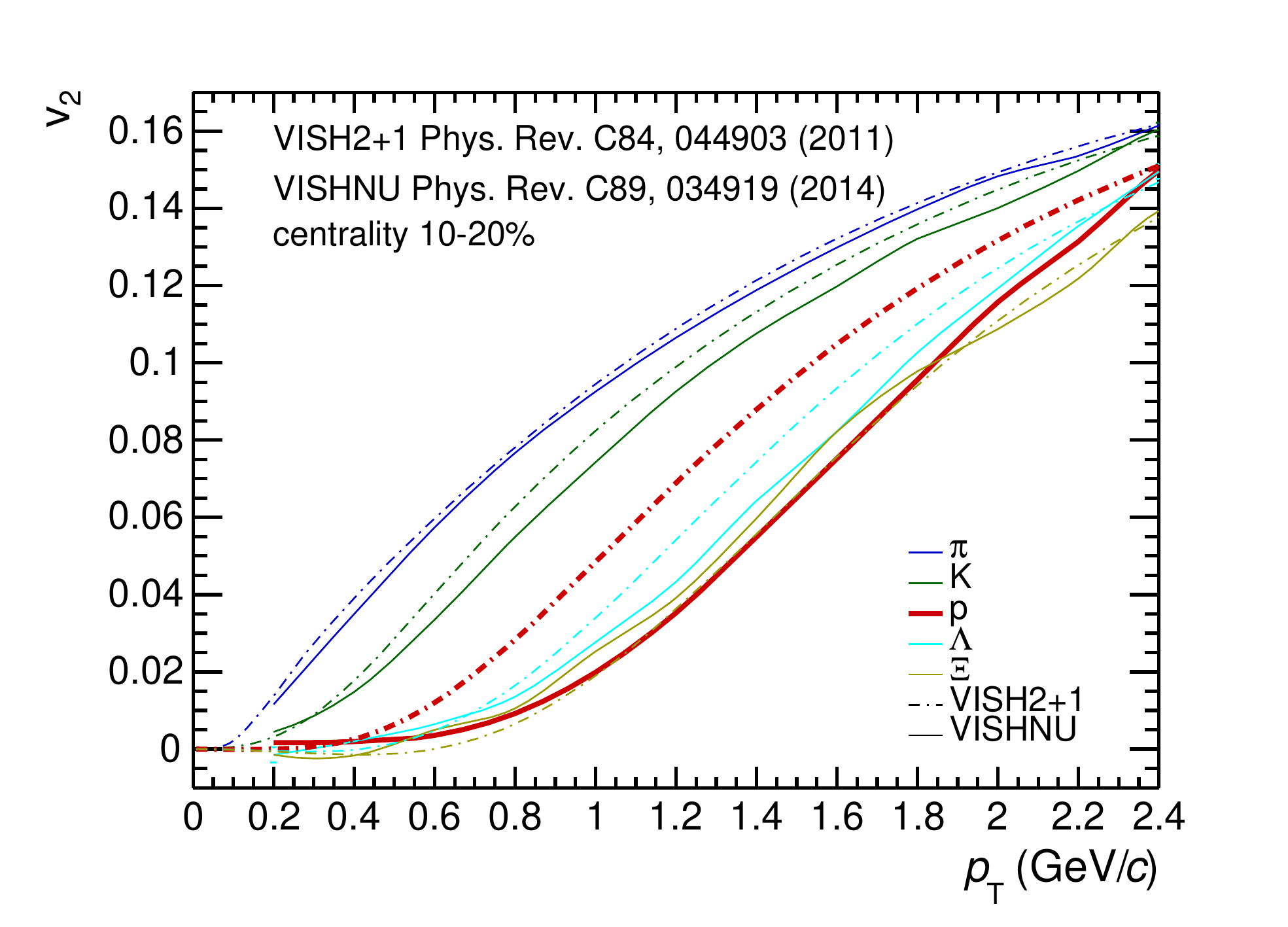}
\includegraphics[width=7.4cm]{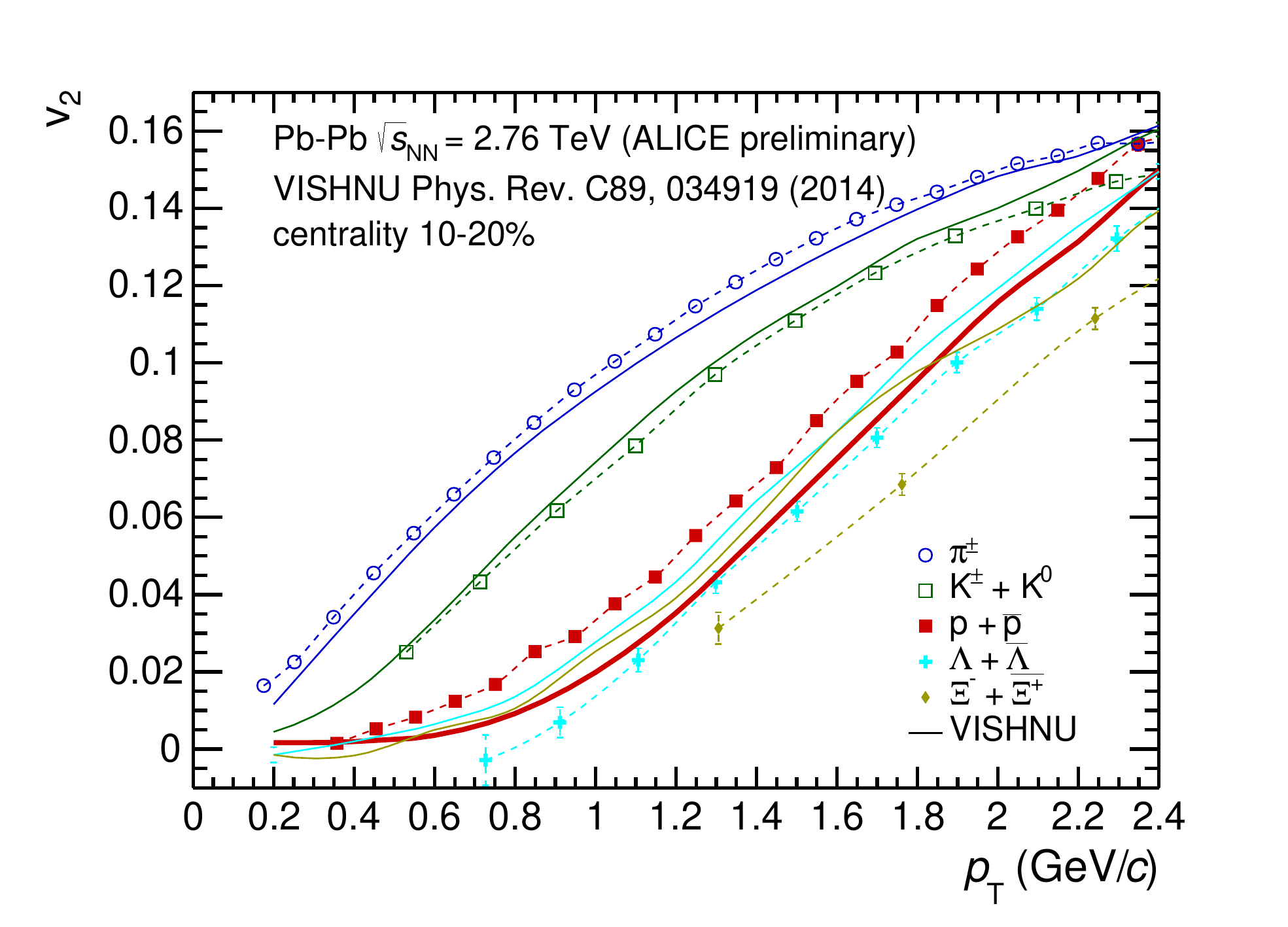}
\caption{
Left: Viscous hydrodynamical calculations ({\sc VISH2+1}) compared to viscous hydrodynamical calculations 
augmented with a hadronic cascade afterburner ({\sc VISHNU}).
Right: {\sc VISHNU} model calculations compared to ALICE measurements~\cite{Abelev:2014pua} 
of $\pt$-differential $v_2$ at $\sqrt{s_{_{\rm NN}}} = 2.76$ TeV for central 10--20\% collisions.
}
\label{fig:figure7} 
\end{figure}
Although there is consensus that much of the dynamical evolution can be modelled by relativistic viscous 
hydrodynamics, the clear deviations between data and the model, such as for the $\pt$-differential $v_2$ of 
baryons in more central collisions, show that not all relevant ingredients are in place.  Augmenting the viscous 
hydrodynamical model calculations with a hadron cascade afterburner is one of the most obvious possibilities 
to improve our understanding and description of the data. 

In the left panel of Fig.~\ref{fig:figure7} viscous hydrodynamical calculations with and without a 
hadronic cascade afterburner, {\sc VISHNU} and {\sc VISH2+1} respectively, are compared. 
The increase in mass splitting between the pions and kaons for {\sc VISHNU} (solid curves) compared to 
{\sc VISH2+1} (dashed curves) illustrates the larger radial flow in the {\sc VISHNU} calculations due to the contribution 
of the hadronic cascade. Between the pions and lambdas a similar increase is observed. However it is seen that the 
mass splitting between the pions and cascades does not change much. This can be understood because 
in most hadronic cascade models a small hadronic re-interaction cross section is assigned to the cascade baryon.  
In a hadron cascade model such as {\sc UrQMD} the magnitude of the hadronic re-interaction cross section is, 
if these cross sections are not measured, related to the number of strange quarks contained in the hadron 
(additive quark model). 
In that case, the amount of radial flow that the hadrons pick up in the hadronic phase depends 
on their strange quark content. 
For the protons the most dramatic change is observed, as they are  
strongly coupled to the flowing matter because of their large interaction cross section with the pions. 
As a result of the different individual hadron-hadron re-interaction cross-sections 
in the hadronic phase the characteristic mass ordering, which was observed in {\sc VISH2+1}, 
is not preserved anymore after including the hadronic cascade in {\sc VISHNU}.  

The comparison between the $\pt$-differential $v_2$ measured by ALICE and the {\sc VISHNU} model is shown in
the right-panel of Fig.~\ref{fig:figure7}. While {\sc VISHNU} gives a better description of the kaons, and 
partially also for the lambdas, it does not improve the description of the cascades. For the protons the modification of 
the $\pt$-differential $v_2$ is so large that, compared to {\sc VISH2+1}, it overcompensates for the original 
difference and again fails to describe the data. In particular the breaking of the mass ordering as seen in {\sc VISHNU} 
is not observed in the data.

Clearly, the current implementations of viscous hydrodynamics coupled to a hadron cascade do not  
describe better the $\pt$-differential $v_2$ of the different particle species in more central collisions. 
Therefore, it is currently still an open question what the origin is of the observed larger radial flow in more central 
collisions. Possible contributions could come from pre-equilibrium flow 
(e.g. AdS/CFT~\cite{vanderSchee:2013pia, Habich:2014jna}),  
contributions from a different baryon production mechanism~\cite{Werner:2012xh,klaus}, or perhaps just  
from improving our knowledge of some of the hadronic re-interaction cross sections. 

\section{$\phi$-Meson Elliptic Flow}

Among the various hadrons measured in heavy-ion collisions, resonances can provide more differential 
information on the hadronic processes because of their different lifetimes and re-interaction cross sections.
An example is the observed non-constant $K^{*0}/K^{-}$ ratio as function of centrality~\cite{Abelev:2014uua}. 
This deviates from thermal and statistical 
model predictions and could be explained by contributions from hadronic re-scattering and regeneration.

Because the amount of hadronic re-scattering and the amount of radial flow picked up in the hadronic phase is 
assumed to depend on the strange quark content the $\phi$-meson spectra and anisotropic flow are 
expected to be sensitive measurements~\cite{Hirano:2007ei,Song:2013qma}. 
The $\phi$-meson contains a strange and anti-strange quark and therefore its $\pt$-differential $v_2$ is expected to be 
primarily generated in the partonic phase. The $\phi$-meson 
$\pt$-differential $v_2$ is expected, in case of mass ordering, to be in-between the protons and lambdas.
However if the contribution to the collective flow depends strongly on the different individual 
hadron-hadron re-interaction cross-sections the mass ordering will be broken 
because the lambdas and in particular the protons will pick up a significant additional amount of flow. 

\begin{figure}[th]
\includegraphics[width=7.4cm]{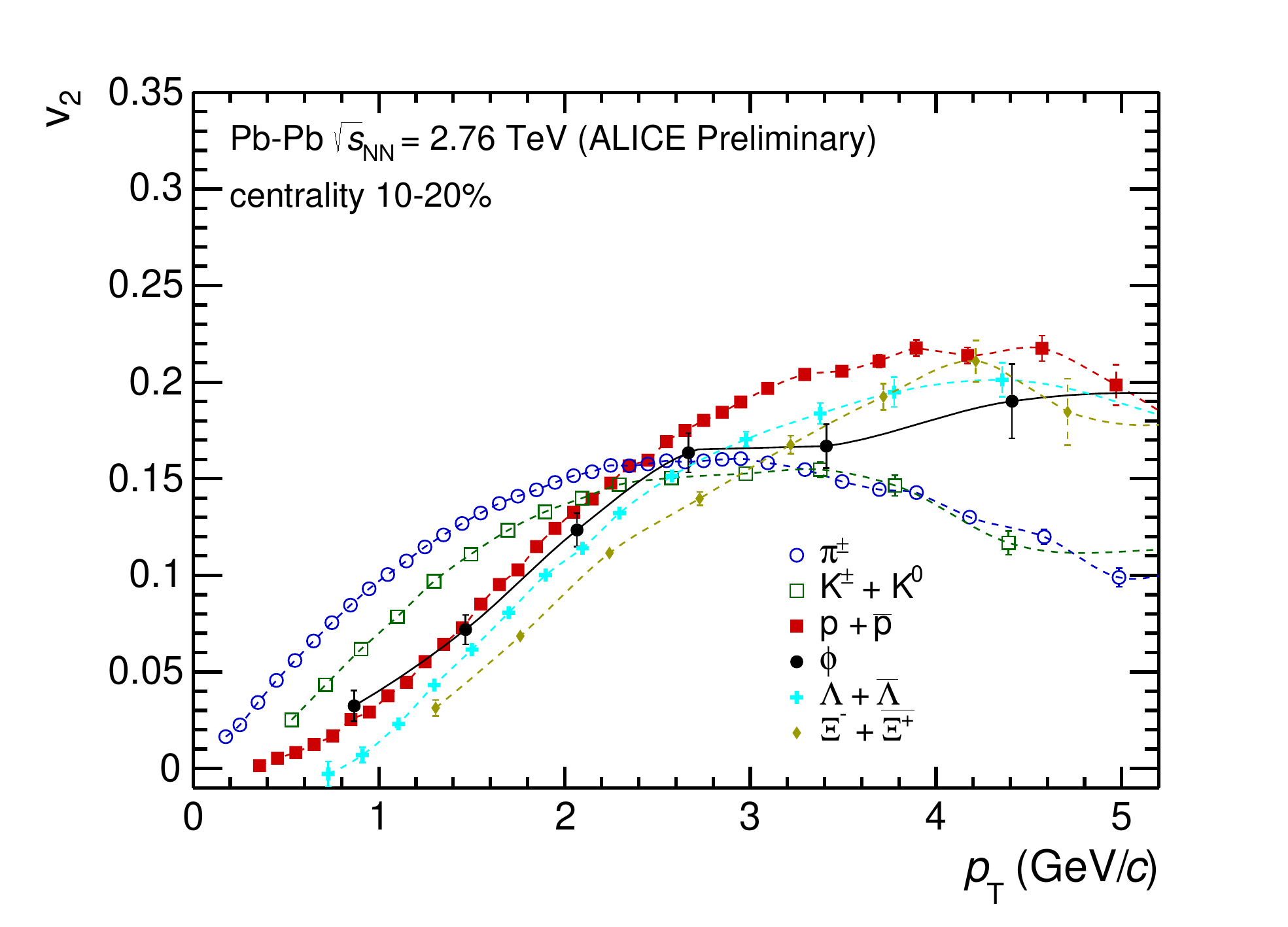}
\includegraphics[width=7.4cm]{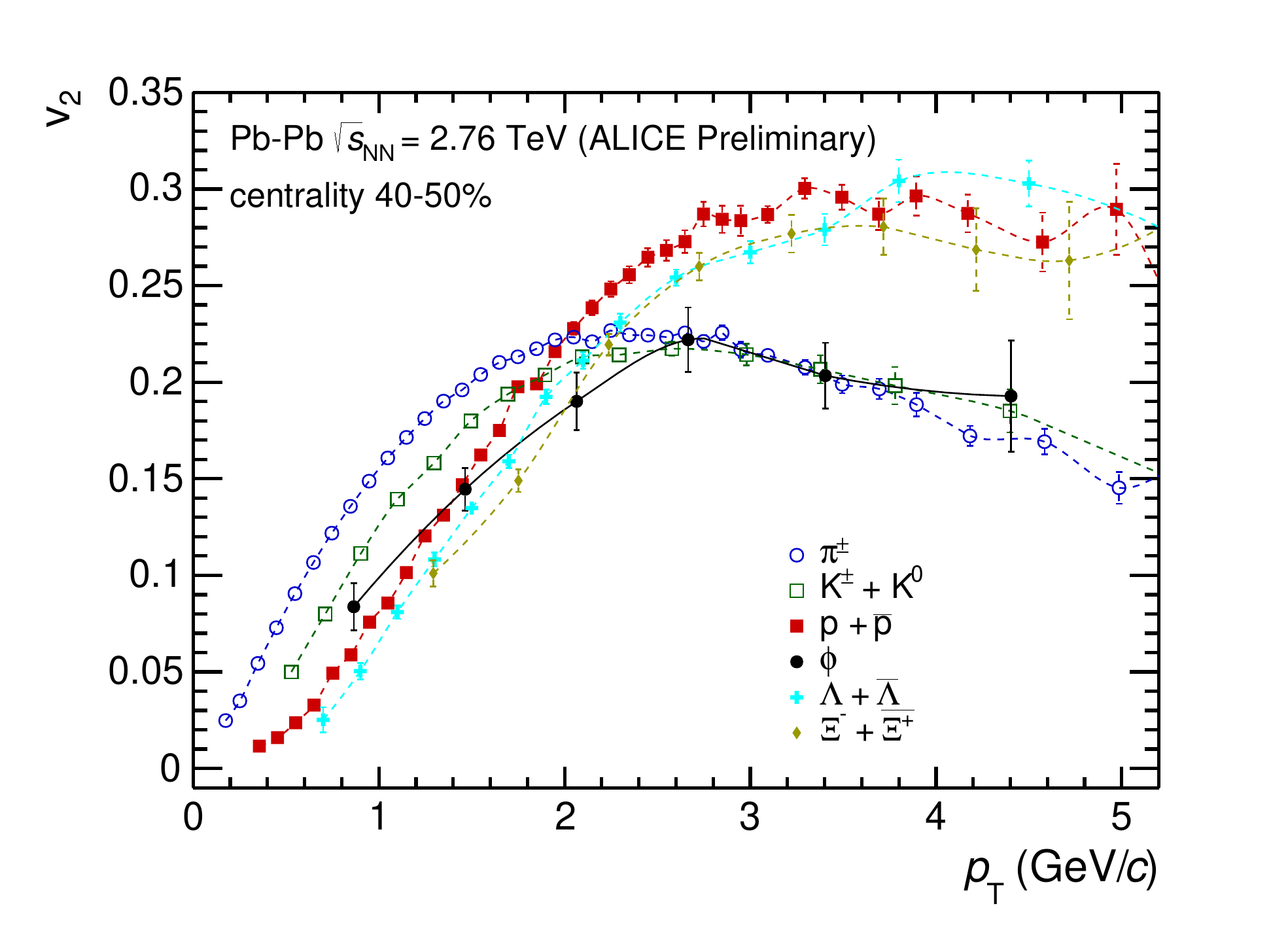}
\caption{
The $\phi$-meson $\pt$-differential $v_2$ compared to other identified particles for two collision 
centralities (from~\cite{Abelev:2014pua}).
}
\label{fig:figure8} 
\end{figure}
Figure~\ref{fig:figure8} shows the $\phi$-meson $\pt$-differential $v_2$ for more central, 10--20\%, and more peripheral, 40--50\%, collisions.
Below $\pt = 2.5$ GeV/$c$ the $\phi$-meson follows within the relatively large uncertainties the mass hierarchy for both centralities. 
However, for the lowest $\pt$ bin there is an indication that the $\phi$-meson $v_2$ is larger than the proton $v_2$. 
Unfortunately, the uncertainties in the ALICE $\phi$-meson $v_2$ measurements are currently still too large at low-$\pt$ to constrain better the 
possible hadronic contribution.

\section{Summary}

In these proceedings I have shown that, with the recent measurements from the LHC and from the 
RHIC beam energy scan, we observe tantalising evidence for  a change in slope of the 
energy dependence of the integrated elliptic flow. From the $\pt$-differential $v_2$ measurements of 
different particle species we see that this can be understood due to increasing collective flow with 
increasing center of mass collision energy. A comparison between the measurements and viscous hydrodynamical 
model calculations with and without a hadronic afterburner shows that currently the contributions from the hadronic phase are not understood yet.  

\section*{Acknowledgements}
I would like to thank the organizers of the conference, in particular Angela Badala for insisting 
that I would prepare this talk and also for the beautiful birthday cake. 
For the contributions to the figures and for the stimulating discussions I am grateful to: 
Joerg Aichelin, Anton Andronic, Tomasz Bold, Panos Christakoglou, Pasi Huovinen, Wojciech Florkowski, 
Ulrich Heinz, Hannah Petersen, Jurgen Schukraft, Alexander Schmah, Chun Shen, Shusu Shi, Huichao Song, 
Jan Steinheimer, Sergei Voloshin, Klaus Werner, Nu Xu, and You Zhou.

\end{document}